\documentclass{llncs}
\usepackage{graphicx} 

\usepackage[backend=bibtex,style=numeric-comp,doi=false,isbn=false,url=false]{biblatex}
\usepackage{fca}

\usepackage{subfigure}

\usepackage{amsmath,amssymb}
\usepackage{mathtools}
\usepackage{colonequals}
\newcommand{\abs}[1]{{\mid}#1{\mid}}
\DeclareMathOperator{\Ext}{Ext}

\usepackage{cleveref}
\usepackage{booktabs,multicol,multirow}
\usepackage{fancybox}
\usepackage{todonotes}

\usepackage{tikz}
\usepackage{tikzscale}
\title{Conceptual Mapping of Controversies}
\author{Claude Draude\inst{2,4}\orcidID{0000-0001-8467-195X}
\and Dominik Dürrschnabel\inst{1,4}\orcidID{0000-0002-0855-4185}
\and Johannes Hirth\inst{1,4}\orcidID{0000-0001-9034-0321} \and Viktoria Horn\inst{2,4}\orcidID{0000-0002-2572-6575} \and Jonathan Kropf\inst{3,4} \and Jörn Lamla\inst{3,4} \and Gerd Stumme\inst{1,4}\orcidID{0000-0002-0570-7908}  \and Markus Uhlmann\inst{3,4} }
\date{March 2024}

\institute{%
 Knowledge \& Data Engineering,
 University of Kassel, Germany\\[0.5ex]
 \email{\{hirth,duerrschnabel,stumme\}@cs.uni-kassel.de}
 \and
Participatory IT Design,
 University of Kassel, Germany\\[0.5ex]
 \email{claude.draude@uni-kassel.de, viktoria.horn@uni-kassel.de}
 \and
 Sociological Theory, University of Kassel, Germany\\[0.5ex]
\email{kropf@uni-kassel.de, lamla@uni-kassel.de, markus.uhlmann@uni-kassel.de}
 \and 
 Interdisciplinary Research Center for Information System Design\\
 University of Kassel, Germany
}

\begin{document}
\maketitle
\begin{abstract}
With our work, we contribute towards a qualitative analysis of the discourse on controversies in online news media. For this, we employ Formal Concept Analysis and the economics of conventions to derive conceptual controversy maps. In our experiments, we analyze two maps from different news journals with methods from ordinal data science. We show how these methods can be used to assess the diversity, complexity and potential bias of controversies.
In addition to that, we discuss how the diagrams of concept lattices can be used to navigate between news articles. 
\end{abstract}
\begin{keywords}
Formal~Concept~Analysis; Economics~and~Sociology~of~Conventions; Conceptual~Scaling; Online Journalism; Recommender Systems
\end{keywords}

\section{Introduction}
Online news platforms have a central role in public opinion making. With their aggregation of perspectives, facts and opinions in the form of news articles, they have a cultivating role. In doing so, they shape the information landscape surrounding controversies. With our work, we introduce a concept-based approach to extract these landscapes in the form of conceptual structures.  
This approach is inspired by research on \emph{mapping controversies} \cite{venturini2012building}.
To qualitatively assess the dimensions of controversies, we employ the sociological theory of \emph{economics of conventions} \cite{young1996economics,diaz2011einfuhrung}, which states that, within a controversy, arguments can be grouped into few \emph{conventions}. 

With the visualization and analysis of concept lattices of
controversies, we address two research problems. First, we show how the concept lattice diagram can be used to navigate between news articles. The current standard for this task is to employ news recommender systems. These, however, are often criticized to overly emphasize economic interests \cite{Bernstein.2021} at the cost of normative public interest. For example, their application is criticized for having a selective and filtering effect~\cite{pariser2011filter,nguyen2014exploring}. This is
particularly critical in the area of news recommendations~\cite{karimi2018news} due to their major influence on public discourse \cite{Helberger.2019}. 
We discuss this in greater detail in the following sections.

The second problem that we address is the analysis of public discourse on a given controversy. For this, we apply methods from ordinal data science to analyze the conceptual controversy maps. These maps do not only provide insights into the complexity of controversies but also allow us to compare different controversies or different perspectives on the same controversy. We demonstrate this by studying the conceptual controversy maps on the controversy of electric mobility for two journals.

\section{Recommender Systems for News Articles}
Due to the increasing amount of online articles and news, some kind of
aggregation or recommendation is required by any reader. One approach to deal with this problem is to employ
news recommender systems (NRS) \cite{karimi2018news} to navigate users
from one article to the next. The field of online journalism comes with high standards that should be respected by an NRS.
There are a multitude of \emph{Ethic Codes} compiled for this field. For instance, the \emph{German Press Code}\footnote{\url{https://www.presserat.de/pressekodex.html}} formulates sixteen criteria with the ultimate goal that \emph{``publishers,
editors and journalists must in their work remain aware
of their responsibility towards the public and their duty
to uphold the prestige of the Press.''}\footnote{\url{https://www.presserat.de/pressekodex.html?file=files/presserat/dokumente/pressekodex/Pressekodex2017english.pdf&cid=218}} (German Press Code).

News recommender systems are often discussed with respect to two types of stakeholders.
The first type of stakeholder has primarily economic interests and includes  platform providers and advertisers. These stakeholders aim at optimizing for user relevance and engagement. For this, they frequently employ supervised machine learning models, such as recommender systems, that solve a ranking problem among articles, e.g., 
\textcite{castells2021novelty}. The quality of these models are mostly evaluated based on prediction accuracy, which reflects the
\emph{relevance} of the recommendation, as well as the
\emph{diversity of content}, \emph{novelty}, and
\emph{serendipity}~\cite{ge2010beyond,karimi2018news}.

The other type of stakeholder pursues public interests which, besides relevance and novelty, value the \emph{journalistic quality}, \emph{privacy of users} and \emph{diversity of perspectives}. Here, the aspect of diversity is quantified with different intentions. For the platform this means that a recommendation of a number of articles should reflect different topics of interest to increase user engagement
\cite{castells2021novelty}. Whereas from a democracy theoretic point of view, a topic should be discussed from multiple perspectives \cite{helberger2018exposure}.

There is a lot of criticism towards current implementations of NRS with
respect to public interests. Recommendations following the principle
of \emph{more of the same}, e.g., \emph{collaborative filtering} or
\emph{content-based filtering} \cite{karimi2018news}, lead to filter
bubbles, polarization, and fragmentation
\cite{pariser2011filter,Helberger.2019}.

Furthermore, privacy concerns may arise by  the inclusion of click data\cite{das2007google}, likes
and dislikes, demographic data \cite{lee2007moners} or users access logs
\cite{lin2012premise,lin2014personalized} in NRS. 
This is further amplified by the (often) black-box character of NRS \cite{diakopoulos2017algorithmic}.

The design of a news recommender system that meets the requirements of both types of stakeholders is a difficult task with, to the best of our knowledge, no currently available solution. With our work, we propose a method for the navigation between news articles that reflects public interests. Instead of recommending articles, we aggregate them into a hierarchical structure, i.e., a
concept lattice. With this approach, we address two problems. First,
the concept lattice diagrams allow users to navigate between
articles. In contrast to recommendations, the user navigates in a
self-determined manner. In \cref{sec:navigation}, we demonstrate how
the line diagrams can be used to navigate to more specialized, general or
complementary articles. Second, the concept lattice structure can be
used to analyze the complexity and dependencies within the public discourse. In
\cref{sec:analysis}, we analyze thoroughly the conceptual structure of articles on the conflict of electric mobility
with methods from ordinal data science
\cite{fca,ordinalDataScience}.
\section{Mapping of Controversies and the Theory of Conventions}

This section outlines the theoretical foundations for an alternative paradigm for navigation through journalistic content. In this regard, \emph{mapping of controversies} \cite{venturini2022controversy} plays a central role, which is rooted in science and technology studies (STS). Research related to STS focuses on controversies that are, for example, caused by technological developments and media evolutions. A particular objective of STS is related to questions on facilitating an appropriate public debate regarding complex controversies. 

In this regard, the mapping of controversies approach aims to represent and visualize the interests, arguments, positions and mutual entanglements of the actors involved in and affected by public controversies. Venturini and Munk describe this approach as follows: ``This effort to unfollow public debate, to care for all viewpoints while not giving everyone the same credit, to explore collective disputes and make them more legible, is a form of mapmaking, although not (or not only) in a geographical or even a graphical sense." \cite[p.\,5]{venturini2022controversy} 
In this paper, we will discuss how Formal Concept Analysis can be used as a method for mapping of controversies, by developing an alternative paradigm for navigating journalistic content.

\begin{table}[t]
  \centering
  \caption{The eight main conventions, their worth, and their
    evaluation criteria from the economics of conventions theory. This
    table is based on work of Diaz-Bone (\cite[Table 4.1]{diaz2018economics} and \cite[Chapter 5]{diaz2011einfuhrung}).}
      \newcommand{\tcell}[1]{\parbox[t][0.9cm][t]{2.3cm}{\baselineskip=0pt #1}}
  \small
  \begin{tabular}{|l||c|c|}
    \hline
    \parbox{2cm}{\;\\Convention}&\parbox{2cm}{\centering Worth/\\Quality}&\parbox{2cm}{\centering Evaluation\\ criteria}\\\hline \hline
    Domestic &\tcell{Tradition,\\handcraft}&\tcell{Esteem,\\reputation}\\\hline
    Market &\tcell{Demand\\ orientation,\\ free exchange}&\tcell{Price}\\\hline
    Industrial&\tcell{Planning and \\ standardization}&\tcell{Efficiency,\\productivity}\\ \hline
    Inspired&\tcell{Grace, \\ nonconformity, \\creativity}&\tcell{Originality,\\innovative\\capacity}\\ \hline
    Opinion&\tcell{Renown\\\;}&\tcell{Amount of \\recognition}\\ \hline
    \tcell{Civic/\\State}&\tcell{Collective\\ interest}&\tcell{Relevance for\\collectivity}\\ \hline
    Green&\tcell{Ecology \\(its integrity)}&\tcell{Environmental\\compatibility}\\ \hline
    Network&\tcell{Activity,\\ self-management}&\tcell{Successful\\projects}\\ \hline
  \end{tabular}
  
\label{fig:conventions}
\end{table}

To this end, we draw on an additional theoretical approach in which controversy plays a central role: the \emph{economics of conventions}, which originated in France \cite{DiazBone.2009}. The starting point in one of the key works of convention economics --- “On Justification” \cite{Boltanski.2007} --- are critical situations in which actors identify problems regarding the coordination of their actions. In such situations, both the criticized and the criticizer are subject to an “imperative of justification” in order to validate their actions or criticism \cite[p. 44]{Boltanski.2011}. Boltanski und Thévenot first identified six principles of justification or so-called “orders of worth”, “social worlds” or “polities”, to which other orders have been added over time \cite[p. 63]{Boltanski.2011} (cf. \cref{fig:conventions}). 

Each order of worth highlights a specific understanding of what is valuable. For example, whereas the civic order of worth focuses on the collective interest and values such as solidarity and equality, the market world is concerned with justifications based on monetary principles of exchange and competition, the opinion polity refers to recognition by others, and the industrial world, in turn, focuses on values of efficiency and productivity \cite[p. 222-286]{Boltanski.2007}. Later developed orders of worth include the project-based world, which prizes activity and successful self-management, and the green world, which emphasizes ecological sustainability \cite{Nachtwey.2024}. 

While orders of justification have different conceptions of what is valuable and just, at the same time they share underlying conceptions of the common good that refer to general interests of humanity. For instance, business companies cannot only refer to their private business interests when they highlight fair competition as a central justification and principle to coordinate interests. Rather, companies must show that fair competition and a market-based justification are relevant for the common good as a whole \cite[p. 108]{Boltanski.2007}.

\section{Case Study: Online journalism about e-mobility}
Online journalism is a predestined field to study the discourse on topics that are relevant to the general public.  Online journalism plays a cultivating role: It collects and aggregates public perspectives and places them in relation to one another. 

One particularly interesting use case is the controversy surrounding electromobility. This is characterized not only by its current relevance, but also by its broad public discourse. The topic of electromobility touches a variety of areas of society. It affects individuals in their choice of transportation, as well as the legislator, who has to create and negotiate a legal framework. It also affects the manufacturers, who are in competition with each other. There is also competition with other means of mobility. Besides these, electromobility is being discussed and studied in terms of its contribution to combating climate change. The extent to which these perspectives are in conflict with each other and what compromises are negotiated can be determined on the basis of news articles.

We analyze this controversy on the basis of two journals of a major German publisher. One of these journals has a general readership, while the second provides information for specific target groups. These allow us to test hypotheses such as 
\begin{center}
    \emph{Does any article that discusses electric mobility from the green convention also discuss it from the industry convention?}
\end{center}

To this end, we created a data set for each journal by working out the various players and perspectives using the economics of conventions for each article. 
For this purpose, we assigned the conventions listed in \Cref{fig:conventions} to the articles. We added, for each journal, new articles until no new combination of conventions was found.\footnote{We may note that we do not claim that the selection is representative for both journals. For future, more extensive investigations we envision the use of attribute exploration \cite{ganter1999attribute} to cover all relevant types of articles. For this study we refrain from doing so due to the big expense in manually classifying articles in terms of their conventions and the limited access to the journal's data source. We can also envision that the use of machine learning can help to speed up the classification process -- with a potential loss of accuracy. Preliminary research in that direction has been conducted \cite{solans2021learning}.}
We ended up with a set of twelve articles of the first journal and fourteen articles of the second journal. 

Each article was assigned by experts to one or more of the conventions of Table~\ref{fig:conventions}, which were refined by additional categories. In this regard, we first distinguished whether conventions are referred to in a positive or negative way. A negative reference is made when conventions are taken up with a critical intention. Here again, we have differentiated whether a convention is criticized internally or externally. The former is the case, for example, when the way in which a market-based regulation is implemented is criticized, but the market convention as such is not problematized. In contrast, external criticism occurs, for example, when the market is criticized from the perspective of the state convention. In the case of a positive reference, we have distinguished whether a convention is negotiated as a justification or as a mere topic. The green world is mobilized as a justification, for example, when ecological sustainability is used as a value in its own right to justify certain decisions for industrial production. If, however, ecological sustainability is pushed into the background of a controversy against the backdrop of justifications of competitiveness or industrial efficiency, but also not completely excluded or criticized, the green convention is mobilized not as a justification, but as a thematic reference. 

The assignments are the result of a careful reflection process by trained sociologists. Ambiguous cases were discussed within the group of experts. 

\section{Conceptual Maps of Controversies}\label{sec:analysis}
With the assignment of the conventions and their refinements as described above, we obtained for each of the two journals a formal context with the news articles as objects. In this section, we first analyze these contexts with methods from Formal Concept Analysis. For this, we assume that the reader is familiar with basic notions of FCA \cite{fca}. Second, we discuss in \Cref{sec:navigation} how concept lattice diagrams can be used to navigate between news articles.

\begin{figure}[t]
  \centering
  \newcommand{\x}{\parbox[c][][c]{1cm}{\large \centering $\times$}}
    \rotatebox{90}{
      \begin{tabular}{|c|c|c|c||l|}
        \hline
        \multicolumn{4}{|c|}{Convention (\textbf{C})}&\\\cline{1-4}
        \multicolumn{2}{|c|}{Justification (\textbf{+})}&\multicolumn{2}{c|}{Critic (\textbf{-})}&\quad \rotatebox{-90}{\Large $\context[S]$}\\ \cline{1-4}
        Justification (\textbf{R})&Thema (\textbf{T})&Intern (\textbf{I})&\multicolumn{1}{c|}{Extern (\textbf{E})}&\\ \hline \hline
        \x&\x&\x&\x&\rotatebox{0}{\textbf{\makebox[8pt]{C}\makebox[20pt]{\;}\makebox[10pt]{}}}\\\hline
        \x&\x&&&\rotatebox{0}{\textbf{\makebox[8pt]{C}\makebox[20pt]{\;}\makebox[10pt]{+}}}\\\hline
        \x&&&&\rotatebox{0}{\textbf{\makebox[8pt]{C}\makebox[20pt]{R}\makebox[10pt]{+}}}\\\hline
        &\x&&&\rotatebox{0}{\textbf{\makebox[8pt]{C}\makebox[20pt]{T}\makebox[10pt]{+}}}\\\hline
        &&\x&\x&\rotatebox{0}{\textbf{\makebox[8pt]{C}\makebox[20pt]{\;}\makebox[10pt]{-}}}\\\hline
        &&\x&&\rotatebox{0}{\textbf{\makebox[8pt]{C}\makebox[20pt]{Int}\makebox[10pt]{-}}}\\\hline
        &&&\x&\rotatebox{0}{\textbf{\makebox[8pt]{C}\makebox[20pt]{Ext}\makebox[10pt]{-}}}\\\hline
      \end{tabular}
  }

    \caption{The scale on which we measure the conventions. \textbf{C} is the placeholder for the specific convention.}
  \label{fig:scale}
\end{figure}

In the sequel, we let the news articles be the objects of a formal context. The attributes of the context and its incidence relation are established with the conceptual scale that is shown in \Cref{fig:scale}. Each convention is scaled with this scale, where \textbf{C} is replaced by the convention at hand. By selecting only some of the conventions, we can narrow or broaden the scope of the analysis. Furthermore, we can fine-tune the conceptual resolution of the scale by either considering all seven attributes, or by dropping the attributes \textbf{C R +}, \textbf{C T +}, \textbf{C Int -}, \textbf{C Ext -}, or by dropping additionally the two attributes \textbf{C} + and \textbf{C} -.

We analyzed twenty-six news articles of two journals on electric mobility. 
For the conventions from \Cref{fig:conventions} we chose the \emph{green},
\emph{state}, \emph{market} and \emph{industrial} convention.
We subdivide the resulting scale attributes into three levels based on the hierarchy 
indicated of the scale values (cf. \Cref{fig:scale}, rows). The first level encodes the 
presence of a convention within an article. The second level adds attributes that 
reflect if the article contains arguments that refer in a \emph{positive} or \emph{critical} sense to a convention. 
The third level distinguishes if the positive reference of a convention can be understood as a \emph{justification} or a mere \emph{topic} and if the criticism of conventions is formulated from an \emph{internal} or a \emph{external} perspective. 
The three levels are defined such that the set of attributes of each level includes the attributes of all preceding 
levels, i.e., $M_{\text{L1}}\subseteq M_{\text{L2}}\subseteq M_{\text{L3}}$.

In the theory of \emph{conceptual views}~\cite{scaling-dimension}, we understand the context of the third level $\context_{\text{L3}}$ as the finest view of the three. In particular, we find that $\Ext(\context_ {\text{L}_{i-1}})\subseteq \Ext(\context_{\text{L}_i})$, and thus
$\context_{\text{L}_{i-1}}$ is a coarser view of $\context_{\text{L}_i}$
\cite{smeasure,smeasure-error}.

\begin{table}[t]
  \centering
  \newcommand{\hcell}[1]{\multicolumn{1}{c|}{\parbox{1cm}{\centering #1}}}
   \caption{The context sizes and the number of concepts for the controversy context of the two journals. The levels represent the conventions (L\,1), valuation +/- (L\,2) and the internalization (L\,3) as depicted in Figure~\ref{fig:scale}.} 
  \begin{tabular}{|c|r||r|r|r|r|r|r|}
    \hline
    \rotatebox{45}{Journal}    &\rotatebox{90}{(Articles)}&\multicolumn{2}{c|}{L\,1 ($\abs{M}{=}4$)}&\multicolumn{2}{c|}{L\,2 ($\abs{M}{=}12$)}&\multicolumn{2}{c|}{L\,3 ($\abs{M}{=}28$)}\\ \cline{3-8}
     &$\abs{G}$&\hcell{$\abs{I}$}&\hcell{$\abs{\BV(\context)}$}&\hcell{$\abs{I}$}&\hcell{$\abs{\BV(\context)}$}&\hcell{$\abs{I}$}&\hcell{$\abs{\BV(\context)}$}\\ \hline
     J\,1&12            &0.75&5&0.55&13&0.41&19\\
    J\,2&14            &0.89&6&0.78&18&0.66&66\\
    J\,1 and J\,2&26&0.82&10&0.67&43&0.54&171\\\hline
  \end{tabular}
  \label{tab:context-stats}
\end{table}

In Table~\ref{tab:context-stats}, we depict the size of each derived context as well as their number of formal concepts. We can infer from the table that both contexts are similarly sized in terms of objects and attributes. The density of all contexts decreases the greater the level of detail is, i.e., from L\,1 to L\,3.
The context of Journal 2 is slightly denser than that of Journal 1.
Thus, the articles of journal two discuss the controversy on average more from more conventions. For the number of concepts, we observe a greater difference between both journals. Their difference is only one at the lowest level. At level three journal two has more than three times as many concepts. Thus, one may argue that the controversy of electric mobility is more complex and diverse discussed by journal two.

\begin{table}[t]
  \centering
  \tabcolsep=2pt
  \caption{The number of occurrences of each convention in articles of both journals.}
      \begin{tabular}{|c|c|c|c||c|c|c|c||c|c|c|c||c|c|c|c||}  
        \hline
        \multicolumn{16}{|c|}{\textbf{Journal\,1}}\\                                                                      
        \multicolumn{4}{|c|}{\textbf{M}: 9}& \multicolumn{4}{|c|}{\textbf{G}: 7}& \multicolumn{4}{|c|}{\textbf{S}: 8}& \multicolumn{4}{|c|}{\textbf{I}: 12}\\\hline
        \multicolumn{2}{|c|}{\textbf{+} 9}&\multicolumn{2}{c|}{\textbf{-} 2}& \multicolumn{2}{|c|}{\textbf{+} 7}&\multicolumn{2}{c|}{\textbf{-} 2}& \multicolumn{2}{|c|}{\textbf{+} 8}&\multicolumn{2}{c|}{\textbf{-} 2}& \multicolumn{2}{|c|}{\textbf{+} 11}&\multicolumn{2}{c|}{\textbf{-} 3}\\ \hline
        \textbf{R}: 9&\textbf{T}: 1&\textbf{I}: 2&\multicolumn{1}{c|}{\textbf{E}: 1}& \textbf{R}: 7&\textbf{T}: 4&\textbf{I}: 2&\multicolumn{1}{c|}{\textbf{E}: 1}& \textbf{R}: 8&\textbf{T}: 3&\textbf{I}: 1&\multicolumn{1}{c|}{\textbf{E}: 1}& \textbf{R}: 11&\textbf{T}: 2&\textbf{I}: 3&\multicolumn{1}{c|}{\textbf{E}: 3}\\ \hline \hline
        \multicolumn{16}{|c|}{\textbf{Journal\,2}}\\                                
        \multicolumn{4}{|c|}{\textbf{M}: 13}& \multicolumn{4}{|c|}{\textbf{G}: 12}& \multicolumn{4}{|c|}{\textbf{S}: 12}& \multicolumn{4}{|c|}{\textbf{I}: 13}\\ \hline
        \multicolumn{2}{|c|}{\textbf{+} 13}&\multicolumn{2}{c|}{\textbf{-} 9}& \multicolumn{2}{|c|}{\textbf{+} 12}&\multicolumn{2}{c|}{\textbf{-} 9}& \multicolumn{2}{|c|}{\textbf{+} 10}&\multicolumn{2}{c|}{\textbf{-} 9}& \multicolumn{2}{|c|}{\textbf{+} 13}&\multicolumn{2}{c|}{\textbf{-} 7}\\ \hline
        \textbf{R}: 13&\textbf{T}: 8&\textbf{I}: 8&\multicolumn{1}{c|}{\textbf{E}: 8}& \textbf{R}: 7&\textbf{T}: 7&\textbf{I}: 8&\multicolumn{1}{c|}{\textbf{E}: 8}& \textbf{R}: 9&\textbf{T}: 1&\textbf{I}: 8&\multicolumn{1}{c|}{\textbf{E}: 8}& \textbf{R}: 12&\textbf{T} 9&\textbf{I}: 7&\multicolumn{1}{c|}{\textbf{E}: 7}\\ \hline
  \end{tabular}
  \label{tab:convention-support}
\end{table}

We complement these findings by a detailed overview (see \cref{tab:convention-support}) on how each convention is supported. First, we observe that journal two supports each convention almost equally. For Journal 1, we observe a greater difference. Similar observations can be made for the second level. Journal 1 has more positive occurrences, whereas Journal 2 is more balanced in that regard. The same applies to the internalization. Overall, Journal 2 reflects a more diverse landscape, whereas Journal 1 is simpler and more one-sided. 

\begin{figure}
\begin{subfigure}{}{}
  \centering
  \colorlet{mivertexcolor}{black!80}
\colorlet{jivertexcolor}{black!80}
\colorlet{vertexcolor}{black!80}
\colorlet{bordercolor}{black!80}
\colorlet{linecolor}{gray}
\tikzset{vertexbase/.style 2 args={semithick, shape=circle, inner sep=2pt, outer sep=0pt, draw=bordercolor},%
  vertex/.style 2 args={vertexbase={#1}{}, fill=vertexcolor!45},%
  mivertex/.style 2 args={vertexbase={#1}{}, fill=mivertexcolor!45},%
  jivertex/.style 2 args={vertexbase={#1}{}, fill=jivertexcolor!45},%
  divertex/.style 2 args={vertexbase={#1}{}, top color=mivertexcolor!45, bottom color=jivertexcolor!45},%
  conn/.style={-, thick, color=linecolor}%
}
\begin{tikzpicture}[scale=1]
  \begin{scope} 
    \begin{scope} 
      \foreach \nodename/\nodetype/\param/\xpos/\ypos in {%
        0/vertex//0/0,
        1/jivertex//-1/1,
        2/jivertex//1/1,
        3/mivertex//0/2,
        4/divertex//-2/2,
        5/vertex//-1/3,
        6/jivertex//4/4,
        7/divertex//5/5,
        8/divertex//-2/4,
        9/divertex//1/5,
        10/divertex//-1/5,
        11/mivertex//0/6,
        12/vertex//1/7
      } \node[\nodetype={\param}{}] (\nodename) at (\xpos, \ypos) {};
    \end{scope}
    \begin{scope} 
      \path (11) edge[conn] (12);
      \path (9) edge[conn] (11);
      \path (2) edge[conn] (3);
      \path (2) edge[conn] (6);
      \path (7) edge[conn] (12);
      \path (4) edge[conn] (5);
      \path (3) edge[conn] (5);
      \path (8) edge[conn] (10);
      \path (0) edge[conn] (2);
      \path (0) edge[conn] (1);
      \path (1) edge[conn] (4);
      \path (1) edge[conn] (3);
      \path (10) edge[conn] (11);
      \path (5) edge[conn] (9);
      \path (5) edge[conn] (8);
      \path (6) edge[conn] (11);
      \path (6) edge[conn] (7);
    \end{scope}
    \begin{scope} 
      \foreach \nodename/\labelpos/\labelopts/\labelcontent in {%
        3/above//{Markt -, Grün -},
        4/above//{Staat -},
        7/above//{Industrie -},
        8/above//{Staat, Staat +},
        9/above//{Grün +, Grün},
        10/above//{Markt +, Markt},
        11/above//{Industrie +},
        12/above//{Industrie},
        1/below//{13},
        2/below//{11},
        4/below//{16},
        5/below//{1, 2, 18},
        6/below//{14},
        7/below//{26},
        8/below//{12, 19},
        9/below//{3},
        10/below//{9}
      } \coordinate[label={[\labelopts]\labelpos:{\labelcontent}}](c) at (\nodename);
    \end{scope}
  \end{scope}
\end{tikzpicture}
  \caption{The concept lattice diagram of articles on electric mobility from journal 1.}
\label{fig:lattice-efahrer}
\end{subfigure}
\begin{subfigure}{}{}
  \centering
  \colorlet{mivertexcolor}{black!80} 
\colorlet{jivertexcolor}{black!80}
\colorlet{vertexcolor}{black!80}
\colorlet{bordercolor}{black!80}
\colorlet{linecolor}{gray}
\tikzset{vertexbase/.style 2 args={semithick, shape=circle, inner sep=2pt, outer sep=0pt, draw=bordercolor},%
  vertex/.style 2 args={vertexbase={#1}{}, fill=vertexcolor!45},%
  mivertex/.style 2 args={vertexbase={#1}{}, fill=mivertexcolor!45},%
  jivertex/.style 2 args={vertexbase={#1}{}, fill=jivertexcolor!45},%
  divertex/.style 2 args={vertexbase={#1}{}, top color=mivertexcolor!45, bottom color=jivertexcolor!45},%
  conn/.style={-, thick, color=linecolor}%
}
\begin{tikzpicture}[scale=0.3,font=\footnotesize]
  \begin{scope} 
    \begin{scope} 
      \foreach \nodename/\nodetype/\param/\xpos/\ypos in {%
        0/vertex//0.0/0.0,
        1/jivertex//1.0/5.0,
        2/jivertex//-5.0/7.0,
        3/jivertex//2.0/10.0,
        4/jivertex//8.0/10.0,
        5/jivertex//-8.0/12.0,
        6/vertex//-4.0/12.0,
        7/vertex//9.0/15.0,
        8/vertex//-7.0/17.0,
        9/vertex//3.0/17.0,
        10/mivertex//10.0/20.0,
        11/vertex//-6.0/21.4,
        12/mivertex//0.0/22.0,
        13/mivertex//4.0/22.0,
        14/divertex//-8.2/24.6,
        15/mivertex//1.0/27.0,
        16/mivertex//1.5/31.0,
        17/vertex//-0.5/34.0
      } \node[\nodetype={\param}{}] (\nodename) at (\xpos, \ypos) {};
    \end{scope}
    \begin{scope} 
      \path (5) edge[conn] (12);
      \path (5) edge[conn] (8);
      \path (6) edge[conn] (13);
      \path (6) edge[conn] (8);
      \path (12) edge[conn] (15);
      \path (13) edge[conn] (15);
      \path (7) edge[conn] (13);
      \path (7) edge[conn] (10);
      \path (4) edge[conn] (7);
      \path (4) edge[conn] (9);
      \path (14) edge[conn] (17);
      \path (11) edge[conn] (14);
      \path (11) edge[conn] (16);
      \path (1) edge[conn] (6);
      \path (1) edge[conn] (7);
      \path (1) edge[conn] (3);
      \path (9) edge[conn] (12);
      \path (9) edge[conn] (13);
      \path (3) edge[conn] (11);
      \path (3) edge[conn] (10);
      \path (0) edge[conn] (4);
      \path (0) edge[conn] (1);
      \path (0) edge[conn] (2);
      \path (16) edge[conn] (17);
      \path (15) edge[conn] (16);
      \path (10) edge[conn] (16);
      \path (8) edge[conn] (11);
      \path (8) edge[conn] (15);
      \path (2) edge[conn] (5);
      \path (2) edge[conn] (6);
      \path (2) edge[conn] (9);
    \end{scope}
    \begin{scope} 
      \foreach \nodename/\labelpos/\labelopts/\labelcontent in {%
        1/above//{Industrie -},
        10/above//{Markt -},
        12/above//{Staat +},
        13/above//{$\quad$\parbox{1.5cm}{\baselineskip=0pt Grün -,\\ Staat -}},
        14/above//{Industrie +, Industrie},
        15/above//{Staat, Grün +, Grün},
        16/above//{Markt +, Markt},
        0/below//{24, 4, 25, 17, 5},
        1/below//{22, 8},
        2/below//{21},
        3/below//{20},
        4/below//{6},
        5/below//{7, 15, 10},
        14/below//{23}
      } \coordinate[label={[\labelopts]\labelpos:{\labelcontent}}](c) at (\nodename);
    \end{scope}
  \end{scope}
\end{tikzpicture}
  \caption{The concept lattice diagram of articles on electric mobility from journal 2.}
      \label{fig:lattice-general}   
\end{subfigure}
\end{figure}

Next, we analyze the structure of their concept lattices. For this, we depict in \Cref{fig:lattice-efahrer,fig:lattice-general} the
concept lattices  for both journals at level 2. First, we observe that the structure of Journal 1 is simpler than that of Journal 2. This can  for example be measured by the order dimension \cite{fca} which is 2 for the first journal and 3 for the second. Also, the second concept lattice has a larger width, i.e., size of the longest anti-chain, of four (J\,2) compared to three (J\,1). In contrast to that is the depth, i.e., size of the longest chain, of eight (J\,1) compared to seven (J\,2). 

In terms of  article annotations, we observe for Journal 2 that a majority of articles are assigned to concepts in the lower part of the concept lattice --- which means that each of them addresses a large variety of conventions. In particular, there are five articles that portrait all conventions from the positive and negative side. Overall, this reflects a very diverse discussion on the electric mobility topic in Journal 2. This is in contrast to Journal 1. Here, almost all concepts --- and in particular those further at the top of the lattice --- have article annotations, and no article is addressing all conventions.

In terms of convention annotations, we observe the overall dominance of the industry convention for Journal 1. Here, every article is in incidence with either the positive or the negative industry convention. This is not the case for Journal 2. Since both contexts share the same set of attributes, i.e., study the same conventions, we can compare their set of intents. We find that, besides the empty and full set of attributes, both contexts share only two intents. These are the sets $\{\emph{Industry +}, \emph{Industry}\}$ and $\{\emph{Industry +, Industry, Market, Market +}\}$. Thus, the journals are very dissimilar with respect to closed sets, i.e., represent very different conceptual controversy maps.

In terms of implications, we find more commonalities.
For example, in both contexts the three implications $\{\emph{Market -}\}\to\{\emph{Market +}\}$,  $\{\emph{State -}\}\to\{\emph{State +}\}$ and $\{\emph{Green -}\}\to\{\emph{Green +}\}$ hold. The second context is also a model of the implication $\{\emph{Industry -}\}\to\{\emph{Industry +}\}$. Thus, both journals exhibit a structural bias towards positive conventions. In Table~\ref{table-bases}, we compiled their canonical bases at Level 1. These support our findings that the industry convention is dominant in Journal 1 and that both journals provide different views on the controversies around electromobility.

\begin{table}[t]
\caption{Canonical bases at level 1 for the two journals.}
\label{table-bases}
\begin{minipage}[h][2cm][t]{0.49\linewidth}
\begin{center}
Journal 1
\end{center}\small
\begin{itemize}
\item[] $\emptyset$ $\to$ Industry
\item[] State, Industry $\to$ Market
\item[] Market, Green, Industry $\to$ State
\end{itemize}
\end{minipage}
\begin{minipage}[h][2cm][t]{0.49\linewidth}
\begin{center}
Journal 2
\end{center}\small
\begin{itemize}
\item[] Green$\to$ Market, State
\item[] State $\to$ Market, Green
\end{itemize}
\end{minipage}

\end{table}

\begin{figure}[!b]
    \textbf{Journal 1:}
    \scriptsize
    \begin{itemize}
        \item[$F_1\colon$] Industry $\enspace >\enspace$ Industry +, Industry + R $\enspace >\enspace$ Market, Market +, Market + R $\enspace >\enspace$ State, \mbox{State +}, State + R $\enspace >\enspace$ Green +, Green, Green + R $\enspace >\enspace$ Green + T $\enspace >\enspace$ Market -, Green - I, Green -, \mbox{Market - I $\enspace >\enspace$} State + T, Green - E, Industry - I, Industry - E, Market - E, Industry - $\enspace >\enspace$ Market + T, State - E, State - I, Industry + T, State -
        \item[$F_2\colon$] Industry $\enspace >\enspace$ Industry +, Industry + R $\enspace >\enspace$ Green +, Green, Green + R $\enspace >\enspace$ Market +, \mbox{Market + R}, Market, State, State + R, State + $\enspace >\enspace$ Green + T $\enspace >\enspace$ Industry + T $\enspace >\enspace$ \mbox{Market -}, Green - I, State - I, Market - I, Green -, State - $\enspace >\enspace$ State + T, Green - E, \mbox{Market + T}, \mbox{State - E}, Industry - I, Industry - E, Market - E, Industry -
        \item[$F_3\colon$] Industry $\enspace >\enspace$ Industry - I, Industry -, Industry - E $\enspace >\enspace$ Industry +, Industry + R $\enspace >\enspace$ \mbox{Market -}, Green - I, State + T, Green + R, Green - E, Green + T, Market +, Market + R, \mbox{Market - I}, Market, State, Market - E, State + R, Green -, Green +, Green, State + $\enspace >\enspace$ Market + T, \mbox{State - E}, State - I, Industry + T, State -
        \item[$F_4\colon$]  Industry $\enspace >\enspace$ Industry +, Industry + R $\enspace >\enspace$ Green +, Green, Green + R $\enspace >\enspace$ Market +, \mbox{Market + R}, Market, State, State + R, State + $\enspace >\enspace$ Green + T $\enspace >\enspace$ State + T $\enspace >\enspace$ \mbox{Market -}, Green - I, Green - E, Industry - I, Industry - E, Market - I, Market - E, Industry -, \mbox{Green - $\enspace >\enspace$} Market + T, State - E, State - I, Industry + T, State -
        \item[$F_5\colon$] Industry $\enspace >\enspace$ Industry +, Industry + R $\enspace >\enspace$ Green +, Green, Green + R $\enspace >\enspace$ Market +, \mbox{Market + R}, Market, State, State + R, State + $\enspace >\enspace$ State - $\enspace >\enspace$ State - E $\enspace >\enspace$ Market -, \mbox{Green - I}, State + T, Green - E, Market + T, Green + T, State - I, Industry - I, Industry - E, Market - I, Market - E, Industry + T, Industry -, Green -
        \item[$F_6\colon$]  Industry $\enspace >\enspace$ Industry +, Industry + R $\enspace >\enspace$ Market, Market +, Market + R $\enspace >\enspace$ State, \mbox{State +}, State + R $\enspace >\enspace$ Market + T $\enspace >\enspace$ Market -, Green - I, State + T, Green + R, \mbox{Green - E}, \mbox{Green + T}, State - E, State - I, Industry - I, Industry - E, Market - I, Market - E, \mbox{Industry + T}, \mbox{Industry -}, Green -, State -, Green +, Green
    \end{itemize}

    \normalsize
    \textbf{Journal 2:}
    \scriptsize
    \begin{itemize}
        \item[$F_1\colon$]Market, Market +, Market + R $\enspace >\enspace$ State, Green, Green + $\enspace >\enspace$ Industry +, \mbox{Industry}, \mbox{Industry + R} $\enspace >\enspace$ State -, State - E, Green - E, Green - $\enspace >\enspace$ Market -, Green - I, \mbox{State - I}, \mbox{Industry - I}, Industry - E, Market - I, Market - E, Industry - $\enspace >\enspace$ Industry + T, \mbox{Market + T $\enspace >\enspace$} State + $\enspace >\enspace$ State + R $\enspace >\enspace$ Green + R $\enspace >\enspace$ Green + T, State + T
        \item[$F_2\colon$] Industry +, Industry $\enspace >\enspace$ Industry + R $\enspace >\enspace$ Market +, Market + R, Market, State, \mbox{Green +}, \mbox{Green $\enspace >\enspace$} State + $\enspace >\enspace$ State + R $\enspace >\enspace$ Green + R $\enspace >\enspace$ Industry + T, \mbox{Market + T} $\enspace >\enspace$ \mbox{Green + T $\enspace >\enspace$} Market -, Green - I, State + T, Green - E, State - E, State - I, Industry - I, Industry - E, Market - I, Market - E, Industry -, Green -, State -
        \item[$F_3\colon$]  Market, Market +, Market + R $\enspace >\enspace$ \mbox{Market -} $\enspace >\enspace$ Green - I, State - I, \mbox{Market - I}, State, Green -, State -, Green +, Green $\enspace >\enspace$ Green + T $\enspace >\enspace$ State + $\enspace >\enspace$ Industry +, Green - E, \mbox{Industry + R}, State - E, Industry - I, Industry - E, Market - E, Industry, Industry - $\enspace >\enspace$ Industry + T, \mbox{Market + T} $\enspace >\enspace$ State + T $\enspace >\enspace$ Green + R, State + R
        \item[$F_4\colon$]  Industry +, Industry $\enspace >\enspace$ Market, Market +, Market + R $\enspace >\enspace$ Industry + T $\enspace >\enspace$ \mbox{Market -}, Market - E $\enspace >\enspace$ Green - I, Green - E, Industry + R, Market + T, State - E, State - I, \mbox{Industry - I}, Industry - E, Market - I, State, Industry -, Green -, State -, Green +, Green $\enspace >\enspace$ \mbox{Green + R $\enspace >\enspace$} State +, State + R $\enspace >\enspace$ Green + T, State + T
        \item[$F_5\colon$] Market, Market +, Market + R $\enspace >\enspace$ State, Green, Green + $\enspace >\enspace$ State + $\enspace >\enspace$ State + R $\enspace >\enspace$ Industry +, Industry, Industry + R $\enspace >\enspace$ Green + T $\enspace >\enspace$ Industry + T, Market + T $\enspace >\enspace$ \mbox{Green + R} $\enspace >\enspace$ Market -, Green - I, State + T, Green - E, State - E, State - I, Industry - I, Industry - E, Market - I, Market - E, Industry -, Green -, State -
    \end{itemize}
    \caption{Two complete greedy ordinal factorizations of the data sets analyzed in this work.
    In both cases, the first ordinal factor $F_1$ is the determinative factor and indicates the  focus of the journal.}
    \label{fig:enter-label}
\end{figure}

\section{Dimensionality and Factor Analysis}
In this section, we extend our analysis of the two journals with ordinal factor analysis. This theory allows us to identify (ordinal) dimensions in their conceptual structures. Each factor is represented as a context of an ordinal scale and can be interpreted as a focus direction within the concept lattices. The decomposition of a concept lattice into a minimal number of such factors is a computationally hard problem. Therefore, we use a greedy algorithm for their computation. We refer the reader for a more thorough discussion of the ordinal factor analysis and its computational aspects to the literature \cite{duerrschnabel_greedy}.

In Figure~\ref{fig:enter-label}, we depict for both journals the result of the ordinal factorization. Each factor $F_i$ is represented by a chain of attributes, where conventions have a higher support in a factor the further to the left they are. 


First, we examine the largest factors of each journal with respect to their incidence support, i.e., the number of article/convention pairs of the concept lattice that can be read from the factor. In Journal 1, the largest factor covers 121 of the 139 incidence pairs (87.05\%) and in Journal 2 the largest factor covers 217 of the 260 incidence pairs (83.46\%). The high supports  suggest that the conventions in both journals follow one main (ordinal) dimension.

Ordinal factors can be described by the attributes that occur first on the ordinal scale, since they have the highest support.
For Journal one, we observe that the main factor follows a sequence of a positive (1) industrial, (2) market, (3) state and (4) then green convention. In comparison, the industry convention occurs later in the main factor of Journal two. Here, the positive market (1) and green (2) conventions are first. Similar to our observations of the last section, we see that the negative conventions are less dominant.

Next, we compare the ordinal factors of both journals quantitatively. To do so, we compare the support of the first factors in the other journal.
If we compute the support of the largest factor of Journal 1 in Journal 2, we find that only 114 of the 260 incidence pairs are covered (43.85\%). The largest factor of Journal 2 in Journal 1 has a support of 63 of the 139 incidence pairs (45.32\%).
Based on this observation, we conclude that the conventions follow a different (main) ordinal dimension in each of the journals.
\section{Navigation from a User's Perspective}
\label{sec:navigation}

In this section, we discuss how the concept lattice diagrams can be used to navigate between news articles. To do so, we first introduce different \emph{purposes of use} for the navigation with the concept lattice. 

For this, assume that a user just read an article of Journal 2, e.g., Article 7. Thus, the user has read positive arguments from the industry, state, green, and market perspective. There are multiple options for the user on how to proceed:
\begin{itemize}
     \setlength{\itemindent}{1.2em} 
\item[\emph{First,}] she can read an article that reflects a more \textbf{specialized or focused} article, i.e., one with fewer conventions. For this, she can choose an article that is annotated at an upper neighbor of Article 7, e.g., Article 23. 
\item[\emph{Second,}] the user can read an article that reflects a \textbf{broader or more general} perspective with more conventions. For this, we choose an article that is annotated at a lower neighbor, e.g., Article 21.
\item[\emph{Third,}] one may be interested in a discussion that is \textbf{different} to what we just read. For this, we chose an article that is annotated to a concept that is incomparable to the one just visited, or one that shares only few conventions with the one just read. One such article might be Article 6. 
\item[\emph{Forth,}] we are able to find articles that are \textbf{complementary} in some conventions. For this, we chose an article that has the negative or positive counterpart to a just visited perspective. For example, Article 20, which is in incidence to the negative market perspective, but shares the positive industry and market perspective. 
\end{itemize}

Another form of navigation employs the meet and join operations of concept lattices. These operations have a special semantic meaning in the studied contexts. An article that is at the meet of two other articles/conventions may find a \textbf{compromise} between two conventions in the given conflict of electric mobility. Similarly, an article that is at their join represents the commonalities between articles. 

From a user-centered perspective, the concept lattice diagram offers the user a more transparent approach to navigate between news articles. Opposite to established navigation paradigms like mentioned in the introduction, users are able to see interconnections like similarity, dissimilarity and compromises between articles and may relate them to a concrete feature based on the content. Usual recommender systems often use opaque mechanisms and non-defined criteria like relevance, popularity and similarity to order possible next articles in lists. 
Criteria may also be designed in a way that they distort the online discourse: For example \citeauthor{robertson2023negativity} found that 
a higher amount of negative words in a headline leads to higher click rates \cite{robertson2023negativity}. Thus, if 'popular' articles are identified by which are the most-clicked articles, this may lead to a disproportionate share of negative news in the top recommendations. 
Moreover, lists as a structure themselves guide users in unconscious ways: The position bias (preferring objects at the top of a list regardless of its actual relevance) was found to be replicated also in recommender systems settings \cite{collins2018position}. The map-like structure of the concept-lattice may overcome such biases and lead to the user to a more reflected way of navigating between articles.

However, this navigation structure based on Formal Concept Analysis is highly demanding for novice users not used to reading concept lattice structures. An initial tutorial or training is needed to make it applicable for lay people. This leads to a possible lower acceptance when using such a navigation form in the context of a highly fast-paced and increasingly incidental \cite{yanardaugouglu2021just, park2021s} use case like online news consumption. 

Another challenge in using such a (visually) more complex navigation par\-a\-digm is that news consumption is becoming mobile-first \cite{wolf2015news}. Forwarded from online search platforms or social media platforms \cite{park2021s, nielsen2014relative}, news consumption is performed a lot on mobile phones reducing the screen size to a minimum and thereby consumption patterns are transferred into a irregular `snacking' of news media \cite{yanardaugouglu2021just} instead of deeply indulging into topics and context information. 

To counteract those challenges, an user-centered and intuitive interactive
version of this form of conceptual mapping of controversies is needed as well as careful selection of use contexts, e.g. specialised journals or news content that is not so timely bound but has general information quality besides actuality. As media users actively and strategically choose media for different purposes in order to meet specific needs and goals \cite{blumler1974uses}, the conceptual mapping of controversies offers a proficient solution for users searching for a representation of multiple perspectives on a specific topic.  
As we have shown, the complexity of this alternative navigational structure may burden the mental load of an individual user and depends on use context, among other things. However, while users may initially find themselves grappling with coming to terms with the navigation, the richness and depth of content accessible through such a structure can ultimately enhance their understanding and engagement. Following this, the alternative navigation makes the diversity of the journalistic landscape visible and accessible and contributes to a more informed public discourse.
\section{Conclusion}
With our work, we contributed towards an evaluation of  public discourse on controversies. To characterize the dimensions of a discourse, we employed the theory of \emph{economic of conventions}. This theory allowed us to measure the dimensions based on few perspectives, called conventions, into which arguments can be grouped. Based on these, we derived conceptual scales which can be applied to any controversy. By deriving concept lattices from these scales, we are able to compute conceptual maps of controversies.

We demonstrated this approach on the controversy regarding electric mobility and analyzed the conceptual controversy maps with state-of-the-art methods from ordinal data science. In our analysis, we derived maps for two journals of a major German news publisher. With our analysis, we were able to show several differences on how this controversy is reflected by these journals. For example, we found that Journal 1 has a simpler conceptual structure and a bias towards specific conventions. On the other hand, for Journal 2, we found that it reflects a more diverse and complex landscape of discourse.

Besides that, we showed how the conceptual controversy maps can be used for navigation between online news articles. For this, we derived four means of navigation. We discussed their potential and limitations based on the perspective of human computer interaction. In addition to that, we compared this approach to the use of recommender systems from a user perspective and normative criteria from the realm of online journalism.


\medskip
\textbf{Funding:}
This work was performed in the project FAIRDIENSTE which was funded by the German Federal Ministry of Education and
Research (BMBF) in its program ``Economical aspects of IT security and privacy'' under grant number
16KIS1249K. 

\printbibliography

\end{document}